\RequirePackage{fix-cm}
\documentclass[smallextended]{svjour3}       
\smartqed  
\usepackage{graphicx}
\begin{document}

\title{Gamma-ray pulsars: What have we learned from ab-initio kinetic simulations?
}

\titlerunning{PIC modeling of gamma-ray pulsars}        

\author{Beno\^it Cerutti}


\institute{B. Cerutti \at
              Univ. Grenoble Alpes, CNRS, IPAG, 38000 Grenoble, France \\
              \email{benoit.cerutti@univ-grenoble-alpes.fr}           
}

\date{Received: date / Accepted: date}

\maketitle

\begin{abstract}
The origin of the pulsed gamma-ray emission in pulsars remains an open issue. The combination of sensitive observations in the GeV domain by AGILE and {\em Fermi}-LAT and increasingly sophisticated numerical simulations have recently brought new insights into our understanding of the pulsed emission and particle acceleration processes in pulsars. Particle-in-cell simulations of pulsar magnetospheres show that the equatorial current sheet forming beyond the light cylinder is the main culprit for magnetic dissipation, particle acceleration and bright high-energy synchrotron radiation all together. The shinning current sheet naturally results in a pulse of light each time the sheet crosses our line of sight, which happens twice in most cases. Synthetic lightcurves present robust features reminiscent of observed gamma-ray pulsars by the {\em Fermi}-LAT and AGILE, opening up new perspectives for direct comparison between simulations and observations.

\end{abstract}

\section{Introduction}

The new generation of gamma-ray space telescopes AGILE and {\em Fermi} have greatly contributed to pulsars. Their number detected in the high-energy gamma-ray band increased from the 6 EGRET pulsars in the late nineties to 117 in the second {\em Fermi}-LAT catalog in 2013 \cite{2013ApJS..208...17A}, becoming the largest number of identified sources in the Galaxy and this number continues to increase with more exposure time and better data-analysis techniques. The main results can be summarized as follow:

(i) Gamma-ray pulsars are all rotation-powered and they can be divided into two separate populations: old, low-field ($B\sim 10^9$G) millisecond pulsars whose rotation period was spun up by accretion and young, high-field ($B\sim 10^{12}$G) isolated pulsars whose rotation period is of order $P\sim 100$ms.

(ii) The gamma-ray luminosity above 100 MeV represents about $\sim1-10\%$ of the total energy budget, i.e., the pulsar spin down. Pulsars are therefore extremely efficient particle accelerators.

(iii) The phase-averaged gamma-ray spectrum is well-modelled by a hard power-law at low energies and an exponential cut-off at a few GeV.

(iv) The pulse profile presents in most cases ($75\%$ chance) two well-separated peaks, sometimes with significant emission in between them (the so-called ``bridge emission'').

(v) Gamma-ray pulses are usually not in phase with the radio pulses suggesting that two distinct regions of the magnetosphere are involved in the emission mechanisms. 

(vi) More pulsars are detected in gamma rays than in radio for a given sensitivity (except for millisecond pulsars where radio emission is systematically observed), suggesting that the gamma-ray beam is wider than the radio beam.

These robust features lead to tight constraints on magnetospheric models of the pulsed emission. For instance, the presence of GeV photons pushes the emission zone away from the star surface to avoid their annihilation by the strong magnetic field, and therefore rules out the polar-cap model. Lightcurve modelling using a vacuum dipolar field (e.g., \cite{1995ApJ...438..314R}\cite{2003ApJ...598.1201D}) or using more realistic force-free fields of an inclined rotator \cite{2010ApJ...715.1282B}\cite{2014ApJ...793...97K} also favour the outer regions of the magnetosphere. Unfortunately, the exact location and, more importantly, the physical mechanisms at the origin of particle acceleration remain out of reach in these models. A self-consistent approach taking into account more physics is thus needed to make further progress.

\begin{table}
\caption{Complete list of pulsar magnetosphere modelling using global PIC simulations in order of publication date. ``GR'' stands for General Relativistic, ``Rad.'' stands for Radiation, and ``Part.'' for Particle.}
\label{tab:1}
\begin{tabular}{llll}
\hline\noalign{\smallskip}
References & Inclination & Part. injection & Extra physics \\
\noalign{\smallskip}\hline\noalign{\smallskip}
Philippov \& Spitkovsky (2014) \cite{2014ApJ...785L..33P} & Aligned & Volume &  \\
Chen \& Beloborodov (2014) \cite{2014ApJ...795L..22C} & Aligned & Pair creation & \\
Cerutti et al. (2015) \cite{2015MNRAS.448..606C} & Aligned & Surface & \\
Belyaev (2015) \cite{2015MNRAS.449.2759B} & Aligned & $\mathbf{E}\cdot\mathbf{B}\neq 0$ & \\
Philippov et al. (2015) \cite{2015ApJ...801L..19P} & Oblique & Pair creation & \\
Philippov et al. (2015) \cite{2015ApJ...815L..19P} & Aligned & Pair creation & GR corrections \\
Cerutti et al. (2016) \cite{2016MNRAS.457.2401C} \cite{2016MNRAS.463L..89C} & Oblique & Surface & Rad. \& polarization \\
Cerutti \& Philippov (2017) \cite{2017A&A...607A.134C} & Oblique & Surface & \\
Philippov \& Spitkovsky (2018) \cite{2018ApJ...855...94P} & Oblique & Pair creation & GR \& radiation\\
Kalapotharakos et al. (2018) \cite{2018ApJ...857...44K} & Oblique & Volume & Radiation \\
Brambilla et al. (2018) \cite{2018ApJ...858...81B} & Oblique & Volume &  \\
\noalign{\smallskip}\hline
\end{tabular}
\end{table}

\section{New insights from global particle-in-cell simulations}

The particle-in-cell (PIC) technique is perfectly suited to this problem. This numerical method inherited from plasma physics models the evolution of an ensemble of charged particles and electromagnetic fields altogether and self-consistently \cite{1991ppcs.book.....B}. In the limit of a large number of particles per Debye length, this technique captures all the physics of a collisionless plasma at a kinetic level, and is particularly well adapted for relativistic and highly magnetized plasmas as found in pulsar magnetospheres. Several studies have already applied this technique to model both an aligned and an oblique magnetosphere (see Table~\ref{tab:1} for an exhaustive list). In all cases, the simulation begins with a non-rotating neutron star with a dipolar magnetic field in vacuum. The challenge is then how to supply the magnetosphere with plasma, this is mostly how these studies differ from one another. Several solutions were proposed, e.g., particle injection from the star surface with a non-zero velocity parallel to the magnetic field \cite{2015MNRAS.448..606C}\cite{2018ApJ...858...81B}, injection in non-ideal MHD regions where $\mathbf{E}\cdot\mathbf{B}\neq0$ \cite{2015MNRAS.449.2759B}, volume injection \cite{2014ApJ...785L..33P}\cite{2018ApJ...857...44K}, or pair creation based on a ad-hoc energy threshold of the particles \cite{2014ApJ...795L..22C}\cite{2015ApJ...801L..19P}.

\begin{figure*}
\centering
\includegraphics[width=\hsize]{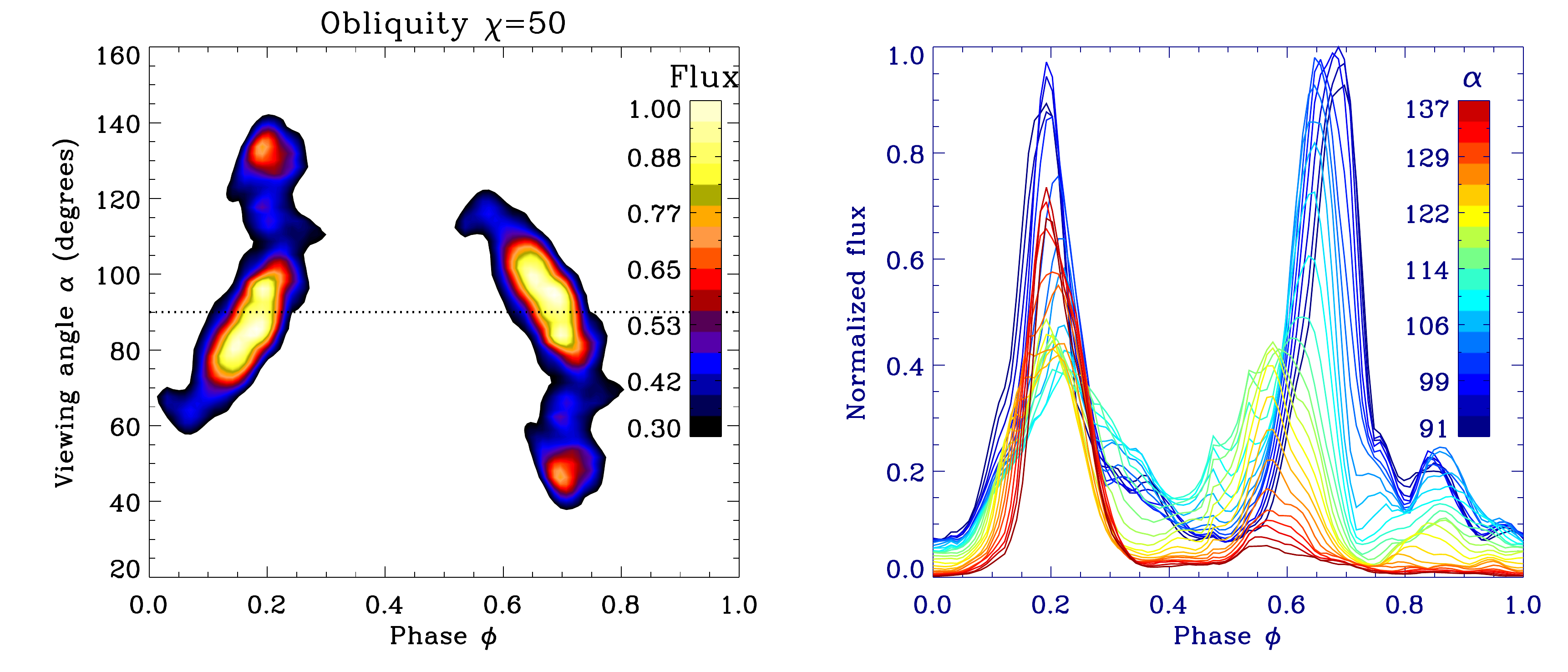}
\caption{{\bf Left}: Gamma-ray emission distribution projected on the sky modelled with global PIC simulations of a pulsar magnetosphere whose magnetic axis makes a $\chi=50^{\rm{o}}$ angle with the rotation axis \cite{2016MNRAS.457.2401C}. The $x$-axis is the normalized pulsar phase $\phi$ and $\alpha$ is the viewing angle with respect to the pulsar rotation axis. Contours show the gamma-ray flux. {\bf Right}: Gamma-ray flux as function of the pulsar phase for a given viewing angle $\alpha$ varying from $90^{\rm o}$ (equator) up to $137^{\rm {\rm o}}$.}
\label{fig_lightcurves}
\end{figure*}

Although different in detail, these studies all agree that the equatorial current sheet forming beyond the light-cylinder radius ($R_{\rm LC}\equiv cP/2\pi$) plays a major role in accelerating particles as anticipated by \cite{1996A&A...311..172L}. Relativistic reconnection dissipates a significant fraction of the pulsar spindown ($\sim 10-20\%$ within $2R_{\rm LC}$) which is then channelled efficiently into relativistic particles and energetic synchrotron radiation \cite{2016MNRAS.457.2401C}\cite{2018ApJ...855...94P} (however, see \cite{2018ApJ...857...44K}). Figure~\ref{fig_lightcurves} shows the imprint on the sky of the gamma-ray emission pattern (``skymap'') directly deduced from PIC simulations under the optically thin approximation for a pulsar with a magnetic moment inclined at a $\chi=50^{\rm o}$ obliquity angle with respect to the rotation axis. The skymap presents two bright features concentrated in the equatorial regions. If the line of sight crosses the equator, an observer will see two bright and symmetric pulses of light separated by roughly half a pulsar rotation period. At intermediate latitudes (viewing angle $\alpha\sim 100^{\rm o}-120^{\rm o}$), the distance between the two peaks shrinks and a significant bridge component appears. At higher latitudes ($\alpha>120^{\rm o}$), the lightcurve becomes more and more asymmetric until we are left with just a single pulse of light at $\alpha\sim 140^{\rm o}$. 

More generally, the shape of the two bright features in the skymaps is set by the geometry of the current sheet. A pulse of light is seen each time the observer's line of sight crosses the sheet. In most cases, this happens twice by period and therefore this provides a natural explanation for the observed {\em Fermi} lightcurves. This scenario also fits well with a wide gamma-ray beam in the equator that is misaligned with a thinner radio beam originating from the polar cap as suggested by observations. While the comparison with gamma-ray observations has remained rather qualitative so far, it is already possible to perform lightcurve fitting. Preliminary results applied to the second {\em Fermi}-LAT pulsar catalog using a chi-squared method show three interesting features: (i) millisecond pulsars tend to be more aligned than young isolated pulsars, (ii) there is a hint of an alignment of the magnetic axis on a $10^5-10^6$yrs timescale which is consistent with what is seen in radio \cite{2010MNRAS.402.1317Y} and with the theoretical prediction proposed in reference \cite{2014MNRAS.441.1879P}, and (iii) the magnetic axis is nearly randomly distributed for very young pulsars, suggesting that there may be no preferential orientation at birth (Alo\"is de Valon, private communications).

Another powerful diagnostic to test even further the scenario depicted by PIC simulations is polarization. This observable is not yet accessible with current gamma-ray telescopes (see, however, reference \cite{2017AIPC.1792g0022G}) but it can be readily inferred from PIC simulations as in reference \cite{2016MNRAS.463L..89C}. The measurement of polarization will be the ultimate way to disentangle between models because it is highly sensitive to the magnetic field geometry within the emitting regions. PIC simulations predict that each pulse of light should be accompanied by a $180^{\rm o}$ swing of the polarization angle. This is the consequence of the passage of the line-of-sight through the current sheet where the magnetic field geometry (mostly toroidal) quickly flips from one polarity to the other over the duration of the pulse. The predicted degree of linear polarization is of order $15\%$ on-pulse and $30\%$ off-pulse in the gamma-ray band. Phase-resolved linear polarization of the incoherent pulsed emission (as opposed to the coherent emission in the radio band) in optical was only measured in the Crab pulsar \cite{2009MNRAS.397..103S}. The data show clear sign of a swing of the polarization angle at each pulse of light. Motivated by the fact that the optical lightcurve is almost identical to the gamma-ray lightcurve, comparison with simulations of the gamma-ray polarization is tantalizing. In reference \cite{2016MNRAS.463L..89C}, the polarization properties can be explained by a unique solution given by a pulsar obliquity of order $\chi=60^{\rm o}$ and viewing angle $\alpha=130^{\rm o}$, which is consistent with independent estimates based on the shape of the Crab Nebula in X-rays \cite{2012ApJ...746...41W}.

\section{Conclusion}

Sensitive gamma-ray observations combined with increasingly sophisticated numerical simulations lead to rapid and major advances in our understanding of pulsar magnetospheres. State-of-the-art PIC simulations allow for the first time to probe some of the most complex magnetic dissipation and particle acceleration processes within the equatorial current sheet forming beyond the light cylinder. Future studies should overcome some of the severe limitations of the PIC simulations which are restricted to unrealistic small-scale separations. A possible solution to circumvent these difficulties may be to develop new hybrid methods, for instance combining the force-free electrodynamic approach for the bulk of the flow and the PIC approach for the microscopic dissipative regions at the polar caps and within the current sheet. This is a promising way to scale simulations up to realistic parameters and to better resolve the polar cap discharge physics and ultimately to understand the origin of the radio emission.

\begin{acknowledgements}
I would like to thank the organisers for their kind invitation and hospitality during the symposium. This work was supported by CNES and Labex OSUG@2020 (ANR10 LABX56).
\end{acknowledgements}

\bibliographystyle{spphys}       
\bibliography{agile2017}   

\end{document}